\documentclass[pdflatex,sn-mathphys-num,iicol]{sn-jnl}
\usepackage{graphicx}
\usepackage{amsmath}
\usepackage{amssymb}
\usepackage{algorithm}
\usepackage{placeins}
\usepackage{algpseudocode}
\usepackage{float}
\usepackage{amsmath,amssymb,bm}
\usepackage{booktabs}
\usepackage{multirow}
\usepackage{array}
\usepackage{braket}
\usepackage{tabularx}
\usepackage{siunitx}
\usepackage{tikz}
\usepackage{natbib}
\usepackage{graphicx}
\graphicspath{{figures/}}
\usepackage{bm}
\usetikzlibrary{arrows.meta, positioning, shapes.geometric, fit, calc}
\newcommand{\insertfigure}[2]{%
  \IfFileExists{#1}{%
    \includegraphics[width=#2]{#1}%
  }{%
    \fbox{%
      \parbox[c][4.2cm][c]{0.94\linewidth}{%
        \centering
        \textbf{Figure placeholder}\\[0.5em]
        \texttt{\detokenize{#1}}
      }%
    }%
  }%
}

\begin{document}
\title[Classical and quantum learning for Higgs classification]
{Sample-Size Scalability and Representation Sensitivity of Quantum Machine Learning Models for Higgs Boson Event Classification}

\author*[1]{\fnm{Shiksha} \sur{Bhardwaj}}
\email{shikshabhardwaj87@gmail.com}
 \author[1]{\fnm{Ankhi} \sur{Roy}}\email{ankhi@iiti.ac.in}

\affil*[1]{%
  \orgdiv{Department of Physics},
  \orgname{Indian Institute of Technology Indore},
  \orgaddress{
    \city{Simrol},
    \state{Madhya Pradesh},
    \postcode{453552},
    \country{India}%
  }
}



\abstract{

Quantum machine learning offers new approaches to multivariate event classification in high-energy physics,
but its performance relative to optimized classical methods remains uncertain under limited training data and restricted quantum resources.
We investigate the sample-size scalability and representation sensitivity of a quantum support vector machine and a variational
quantum classifier for Higgs-boson signal--background discrimination using simulated events
from the ATLAS Higgs Machine Learning Challenge. The models are trained on balanced subsets
ranging from 250 to 4000 events and evaluated on held-out samples preserving the original class distribution. 
Principal component analysis (PCA) and autoencoder (AE) representations are examined to reduce the input dimensionality and quantum-resource requirements.
Four optimized classical classifiers are used as benchmarks under a common preprocessing, validation, and evaluation protocol. 
We assess training-sample scalability in classical and quantum classifiers using a four-component PCA representation and examine representation sensitivity through a 
secondary matched-event comparison of four- and eight-component PCA representations.  
Within the evaluated Higgs-boson classification setting, the QSVM showed stronger performance than the VQC at larger training sample sizes, 
while remaining below the best-performing classical classifiers.

}

\maketitle


\section{Introduction}
\label{sec:introduction}
The discovery of the Higgs boson by the ATLAS and CMS experiments at the
Large Hadron Collider (LHC) established a central component of the Standard
Model of particle physics~\cite{atlas2012observation,cms2012higgs}. Precision measurements of Higgs-boson
production and decay provide stringent tests of the Standard Model and
sensitive probes of possible physics beyond it~\cite{higgsPhysicsReview}. These measurements require the identification of
rare signal processes in the presence of substantially larger and often
kinematically similar backgrounds, making Higgs event selection a challenging
multivariate classification problem~\cite{adambourdarios2015higgs,baldi2014searching}.
Machine-learning methods have consequently become important tools in
high-energy-physics analyses because they can exploit correlations among
measured and derived observables that are difficult to capture with
conventional cut-based selections~\cite{baldi2014searching,baldi2015enhanced}. Nonlinear classifiers, including
kernel methods and tree ensembles, have demonstrated strong performance in
collider-event classification~\cite{adambourdarios2015higgs,chen2016xgboost}. 
For physics applications, however, predictive discrimination alone
is not sufficient. Classification metrics such as ROC--AUC assess
the ability of a model to distinguish signal from background across
different decision thresholds. 
In practice, a specific decision threshold is chosen to select
signal-like events, which determines the resulting signal and
background yields. The approximate median significance (AMS)
provides a physics-oriented measure of the expected statistical
significance of the selected signal relative to the background.
We therefore evaluate both classification performance and AMS
to assess the models from predictive and physics-oriented
perspectives. 
Quantum machine learning (QML) provides an alternative framework for
constructing nonlinear classifiers~\cite{biamonte2017qml}. 
Quantum-kernel methods encode classical data into
quantum states and evaluate similarities through state overlaps, whereas
variational quantum classifiers combine parameterized quantum circuits with
classical optimization~\cite{havlicek2019supervised,schuld2019feature,cerezo2021vqa}.
Applications in high-energy physics have explored quantum annealing~\cite{mott2017higgs,terashi2021event},
variational classifiers~\cite{wu2021vqc,blance2021vqc}, and quantum-kernel methods for signal--background
classification~\cite{wu2021qkernel}.
A practical limitation of present gate-based QML models is the restricted
number of input variables that can be encoded with computationally feasible
quantum resources. When one qubit is assigned to each retained feature, the
representation dimension directly determines the qubit requirement and can
also affect circuit complexity and optimization. Dimensionality reduction is
therefore an integral part of the quantum-classification pipeline rather than
a purely technical preprocessing step. Principal-components and learned latent
representations have been investigated for high-energy applications~\cite{odagiu2025representations}.
Despite continuing progress in QML for high-energy physics event classification, the
relative behaviour of quantum-kernel, variational, and optimized classical
models under controlled changes in both training-sample size and
representation dimension remains less well characterized. Many existing
studies primarily demonstrate the feasibility of particular quantum
architectures, feature maps, or reduced representations. A complementary
question is how these classifier families respond to increasing training
statistics under a common evaluation protocol, and whether a higher-dimensional
representation provides consistent benefits across classical and quantum
models.

The present study addresses these questions using the 
ATLAS Higgs Boson
Machine Learning Challenge dataset~\cite{adambourdarios2015higgs}. Four classical classifiers: logistic
regression, an RBF-kernel support vector machine, random forest, and
XGBoost are compared with two gate-based quantum models: a fidelity-kernel
quantum support vector machine (QSVM) and a variational quantum classifier
(VQC). The analysis focuses on two questions:
\begin{enumerate}
    \item How does classifier performance change as the balanced training
    sample is increased from $N=250$ to $N=4000$?
    \item Does increasing the retained PCA representation from four to eight 
    components provide a consistent improvement across classifier
    architectures?
\end{enumerate}

All preprocessing transformations, representation choices, model configurations, 
and operating thresholds are established using training and validation data only and held 
fixed during test evaluation. Representation screening compares principal-component analysis
(PCA) and autoencoder (AE) embeddings with
$d\in\{2,4,6,8\}$, where $d$ denotes the dimensionality
of the reduced feature representation. The comparison is
performed using validation data. PCA provides a linear projection 
onto orthogonal directions ordered by explained variance, whereas the AE learns a nonlinear 
representation through a \(d\)-dimensional bottleneck. PCA-4 is adopted as the common 
quantum-feasible representation for the primary sample-size scalability study, while PCA-4 versus 
PCA-8 is retained for a predefined secondary analysis of representation sensitivity using a 
separate, disjoint held-out test subset.

The emphasis of this work is therefore not the introduction of a new quantum
classifier, but a controlled comparison of classical, quantum-kernel, and
variational models under common preprocessing, matched evaluation conditions,
event-weighted metrics, and independent representation testing. This framework
allows both predictive performance and physics-oriented consequences to be
assessed consistently across the investigated models.
The remainder of this paper is organized as follows.
Section~\ref{sec:dataset-composition} describes the dataset and permanent data
partitions. Section~\ref{sec:pipeline-framework} presents the experimental
framework, preprocessing and representation protocol, classical benchmarks,
quantum models, and evaluation metrics. Section~\ref{sec:results-discussions} presents the
representation screening, primary sample-size scalability results, and
secondary PCA-4 versus PCA-8 analysis. The corresponding physics and
machine-learning implications are discussed in Section~\ref{sec:conclusions}.
\section{Dataset Composition}
\label{sec:dataset-composition}

This study uses the labelled training sample released for 
the 2014 ATLAS
Higgs Boson Machine Learning Challenge, originally hosted on the Kaggle
competition platform~\cite{atlasHiggsOpenData2014}. The analysed sample
contains simulated proton--proton collision events generated
using the ATLAS detector simulation.
Each event is characterized by 30 physics-motivated input variables,
comprising 17 primitive (\texttt{PRI}) and 13 derived (\texttt{DER})
features. The primitive variables describe reconstructed event kinematics,
including the transverse momenta, pseudorapidities, and azimuthal angles of
the charged lepton, hadronically decaying tau, and reconstructed jets,
together with missing transverse momentum and jet multiplicity. The derived
variables are constructed from combinations of these primitive quantities
and encode higher-level kinematic information relevant to the separation of
Higgs-boson signal from background events~\cite{adambourdarios2015higgs}.

In addition to the 30 classifier variables, each event contains an
\texttt{EventId}, an event \texttt{Weight}, and a class \texttt{Label}.
The original labels \(s\) and \(b\), denoting signal and background,
respectively, are mapped to \(y=1\) and \(y=0\) for the present analysis.
The dataset is retained unchanged, with event identities
preserved through \texttt{EventId} during all subsequent processing.
Before preprocessing or model development, the complete \(250\,000\)-event
sample is divided once into permanent stratified training, validation, and
test partitions in proportions of \(70\%\), \(15\%\), and \(15\%\),
respectively. The resulting partitions contain \(175\,000\), \(37\,500\),
and \(37\,500\) events. Event identifiers are verified to be mutually
exclusive across the three partitions, and the same fixed partitions are
used throughout the analysis. Their class compositions are summarized in
Table~\ref{tab:data-split}.
\begin{table}[t]
\caption{Permanent stratified data partitions used throughout the analysis.}
\label{tab:data-split}
\centering
\small
\begin{tabular}{lrrr}
\toprule
Partition & Events & Background (\%) & Signal (\%) \\
\midrule
Training   & \(175\,000\) & \(65.73\) & \(34.27\) \\
Validation & \(37\,500\)  & \(65.73\) & \(34.27\) \\
Test       & \(37\,500\)  & \(65.73\) & \(34.27\) \\
\bottomrule
\end{tabular}
\end{table}

\section{Methodology and Evaluation Framework}
\label{sec:pipeline-framework}

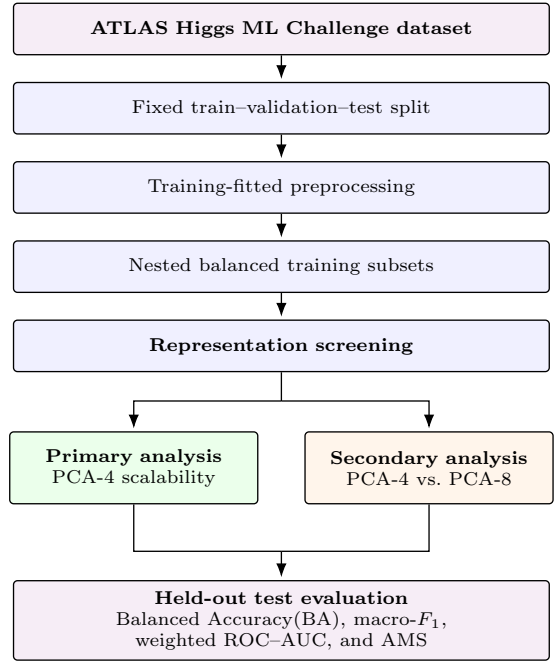
\begin{figure}[t]
\centering

\begin{tikzpicture}[
node distance=3.5mm,
>=Latex,
main/.style={
draw=black,
rounded corners=1.5pt,
line width=0.5pt,
fill=blue!6,
align=center,
text width=6.9cm,
minimum height=6.8mm,
inner sep=2.5pt,
font=\footnotesize
},
primarybox/.style={
draw=black,
rounded corners=1.5pt,
line width=0.5pt,
fill=green!8,
align=center,
text width=3.1cm,
minimum height=9.5mm,
inner sep=2.5pt,
font=\footnotesize
},
secondarybox/.style={
draw=black,
rounded corners=1.5pt,
line width=0.5pt,
fill=orange!8,
align=center,
text width=3.1cm,
minimum height=9.5mm,
inner sep=2.5pt,
font=\footnotesize
},
finalbox/.style={
draw=black,
rounded corners=1.5pt,
line width=0.5pt,
fill=violet!7,
align=center,
text width=6.9cm,
minimum height=10.5mm,
inner sep=2.5pt,
font=\footnotesize
},
flow/.style={
->,
line width=0.5pt,
shorten >=1pt
}
]

\node[main, fill=violet!7] (data)
{\textbf{ATLAS Higgs ML Challenge dataset}};

\node[main, below=of data] (split)
{Fixed train--validation--test split};

\node[main, below=of split] (prep)
{Training-fitted preprocessing};

\node[main, below=of prep] (samples)
{Nested balanced training subsets};

\node[main, below=of samples] (screen)
{\textbf{Representation screening}};

\draw[flow] (data.south) -- (split.north);
\draw[flow] (split.south) -- (prep.north);
\draw[flow] (prep.south) -- (samples.north);
\draw[flow] (samples.south) -- (screen.north);

\coordinate (branch) at ($(screen.south)+(0,-3.8mm)$);

\node[primarybox, anchor=north] (primary)
at ($(branch)+(-1.95cm,-4mm)$)
{\textbf{Primary analysis}\\
PCA-4 scalability};

\node[secondarybox, anchor=north] (secondary)
at ($(branch)+(1.95cm,-4mm)$)
{\textbf{Secondary analysis}\\
PCA-4 vs.\ PCA-8};

\draw[line width=0.5pt] (screen.south) -- (branch);
\draw[flow] (branch) -| (primary.north);
\draw[flow] (branch) -| (secondary.north);

\node[finalbox, anchor=north] (evaluation)
at ($
(primary.south)!0.5!(secondary.south)
+(0,-10mm)
$)
{\textbf{Held-out test evaluation}\\
Balanced Accuracy(BA), macro-$F_{1}$,\\
weighted ROC--AUC, and AMS};

\coordinate (merge) at ($(evaluation.north)+(0,3.8mm)$);

\draw[line width=0.5pt]
(primary.south) |- (merge);

\draw[line width=0.5pt]
(secondary.south) |- (merge);

\draw[flow] (merge) -- (evaluation.north);

\end{tikzpicture}

\caption{
Overview of the experimental workflow. The primary analysis evaluates
sample-size scalability using PCA-4, while the secondary analysis
examines representation sensitivity through a PCA-4 versus PCA-8
comparison. Final performance is evaluated on held-out test data.
}
\label{fig:experimental-workflow}

\end{figure}

The analysis follows a leakage-controlled workflow in which 
preprocessing,
representation learning, model selection, and final testing are kept
separate. Figure~\ref{fig:experimental-workflow} summarizes the complete
protocol. For each training-size and training-seed configuration,
preprocessing transformations, representation models, and classifiers are
fitted using only the corresponding training subset and are subsequently
applied unchanged to validation and test events. Validation data are used for
representation screening, hyperparameter selection, and operating-threshold
determination, while the test data remain excluded from all model-selection
decisions.
For the computationally intensive matched evaluations, stratified random 
sampling is used to construct one fixed validation subset and two fixed, mutually disjoint 
test subsets from their respective permanent partitions. 
Each subset comprises 800 events: 526 background and 274 signal events.
The validation, primary test, and secondary test subsets are generated using 
fixed disjoint sampling random states.
The primary test subset is used for the PCA-4 sample-size scalability study and the 
classical–quantum classifiers, whereas the secondary test subset is reserved exclusively 
for the paired PCA-4 versus PCA-8 analysis. These sampling random states determine only 
the evaluation-event identities and are separate from the training seeds 
used to quantify run-to-run variability.

\subsection{Physics-Oriented Preprocessing and Representation Protocol}
\label{subsec:preprocessing-representation}

Several observables in the original dataset are undefined when the event
does not contain the required reconstructed objects and are represented by
the sentinel value \(-999\)~\cite{pujari2026stability}. These entries are treated as missing values.
For each training size and seed, missing values are imputed using medians
estimated separately within the jet-multiplicity categories
\texttt{PRI\_jet\_num} $\in\{0,1,2,3\}$.
If a category-specific median cannot be determined, the corresponding global
median estimated from the same training subset is used as a fallback.
The resulting imputation statistics are then applied unchanged to the
associated validation and test events.
The jet-multiplicity variable \texttt{PRI\_jet\_num} is retained as an input
feature. In addition, a binary indicator is introduced to record whether
\texttt{DER\_mass\_MMC} was missing in the original event, increasing the
input dimension from 30 to 31 variables. Following imputation, the 30
original physics variables are standardized exclusively from the
corresponding training subset. The binary missing-mass indicator is retained
in its original \(\{0,1\}\) form.
To investigate sample-size scalability, balanced training subsets are
constructed at $N\in\{250,500,1000,2000,4000\}$
for training seeds \(r\in\{42,52,62\}\), with equal numbers of signal and
background events. Within each seed, the subsets are nested,
\begin{equation}
\mathcal{D}_{250}^{(r)}
\subset
\mathcal{D}_{500}^{(r)}
\subset
\mathcal{D}_{1000}^{(r)}
\subset
\mathcal{D}_{2000}^{(r)}
\subset
\mathcal{D}_{4000}^{(r)},
\label{eq:nested-training}
\end{equation}
so that increasing \(N\) adds new events while retaining those already
present at smaller training sizes.
The resulting 31-dimensional event representation is subsequently reduced
before classifier training. This step is particularly relevant to the
quantum models considered here, for which one qubit is assigned to each
retained component. Principal component analysis (PCA)~\cite{jolliffe2002pca}
and a feed-forward autoencoder (AE)~\cite{hinton2006reducing} are therefore
examined at
$d\in\{2,4,6,8\}$.
For every training size and seed, the PCA and AE models are fitted only on
the corresponding balanced training subset and then applied without
refitting to validation and test events.
The AE is optimized using mean-squared
reconstruction error with early stopping based on an internal split of the
training subset, and the parameters corresponding to the lowest internal
validation loss are restored before the encoder is applied externally.
Representation screening is performed with an RBF--SVM used as a common
nonlinear downstream probe. Event-weighted ROC--AUC on the validation data is
used as the primary screening metric, and results are aggregated across the
three training seeds separately for each value of \(N\). The purpose of this
stage is to compare the downstream utility of the reduced representations
under a common classifier rather than to identify a representation that is
universally optimal across model architectures.
Based on this validation-only screening, PCA-4 is fixed before final test
evaluation as the common representation for the primary sample-size
scalability and classical--quantum comparison. Another PCA-4 versus PCA-8 study
is retained for the
predefined secondary representation-sensitivity analysis. The corresponding
screening and test results are presented in
Section~\ref{sec:results-discussions}.

\subsection{Classical and quantum classifiers}
\label{subsec:classical-quantum-models}

The classification study includes four classical benchmarks and two quantum
models. Logistic regression (LR) is used as a linear reference, while the
RBF--SVM~\cite{cortes1995} provides the classical kernel-based counterpart to
the quantum support vector machine. Random forest (RF)~\cite{Breiman2001RandomForests}
and XGBoost (XGB)~\cite{chen2016xgboost} serve as nonlinear tree-ensemble
benchmarks based on bagging and gradient boosting,
respectively. The quantum models comprise a fidelity-kernel QSVM and a
variational quantum classifier (VQC), representing two distinct gate-based
learning paradigms: a fixed quantum-kernel method and a trainable hybrid
quantum--classical model.
All classifiers are evaluated under the common preprocessing,
representation, model-selection, and test protocols described above. Since
the classical methods are well established, their standard algorithmic
formulations are not reproduced here. Instead, the corresponding software
implementations, hyperparameter-search spaces, and validation-selected
configurations are reported to ensure reproducibility. The methodological
discussion therefore focuses on the QSVM and VQC, for which the quantum
encoding, circuit architecture, kernel construction, variational
optimization, and measurement procedures are central to the present study.
The classical implementations use standard scikit-learn interfaces where
applicable~\cite{pedregosa2011scikit}.

\subsection{Quantum models}
\label{subsec:quantum-models}

Quantum approaches to Higgs signal--background classification have been
investigated using quantum kernels, variational classifiers, quantum annealing,
and hybrid quantum--classical methods
~\cite{mott2017higgs,belis2021higgs,wu2021vqc,wu2021qkernel,
fadol2024cepc,belis2024guided,odagiu2025representations,
maulana2026circuit,aithaddou2026hybrid}.
In the present study, two complementary gate-based models are considered:
a fidelity-kernel quantum support vector machine (QSVM) and a variational
quantum classifier (VQC). The QSVM evaluates pairwise similarities in a
quantum feature space, whereas the VQC learns a parameterized quantum
decision function through hybrid quantum--classical optimization
~\cite{havlicek2019supervised,schuld2019feature,cerezo2021vqa}.
Following PCA, each event is represented by
\begin{equation}
\bm{x}=(x_1,\ldots,x_d)\in\mathbb{R}^{d},
\end{equation}
where \(d=4\) for the primary 
analysis and \(d\in\{4,8\}\) for the secondary analysis 
representation-sensitivity study. Before quantum encoding, each PCA component
is rescaled using statistics determined exclusively from the corresponding
training subsets
with validation and test values outside the training range clipped to
\([0,\pi]\). This transformation therefore introduces no information from the
evaluation samples. Since the adopted Qiskit feature maps use
\texttt{feature\_dimension}$=d$, PCA-4 and PCA-8 correspond to four- and
eight-qubit circuits, respectively.

\subsubsection{Quantum support vector machine}
\label{subsec:qsvm}

\begin{figure*}[t]
\centering
\includegraphics[width=0.94\textwidth]{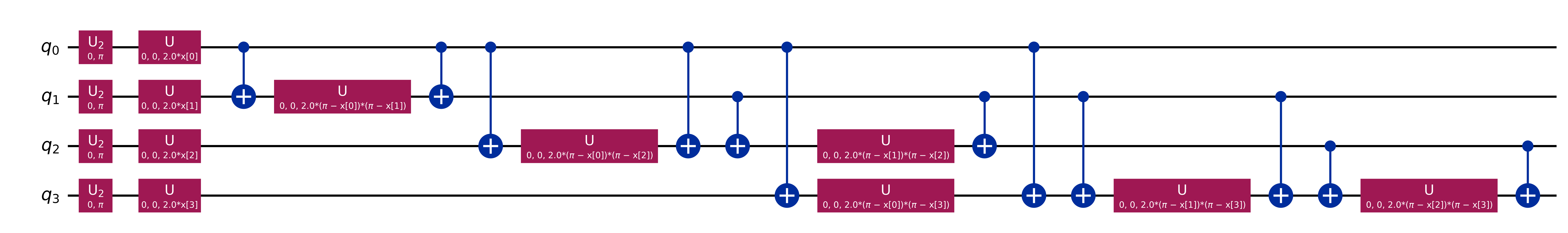}
\caption{
Quantum-kernel construction for the primary PCA-4 QSVM. The four rescaled
PCA components parameterize a four-qubit \texttt{ZZFeatureMap} with one
repetition and full entanglement. Pairwise state fidelities define the
quantum Gram matrix, which is subsequently supplied to a classical support
vector classifier.
}
\label{fig:qsvm-circuit}
\end{figure*}

The QSVM replaces the conventional SVM kernel with the fidelity between
quantum states generated by a data-dependent feature map
~\cite{havlicek2019supervised,schuld2019feature}. For an encoded event
\(\widetilde{\bm{x}}\), the corresponding quantum state is
\begin{equation}
\left|\phi(\widetilde{\bm{x}})\right\rangle
=
U_{\phi}(\widetilde{\bm{x}})
\left|0\right\rangle^{\otimes d},
\label{eq:qsvm-state}
\end{equation}
and the quantum kernel is defined as
\begin{equation}
K_{\mathrm Q}
(\widetilde{\bm{x}}_i,\widetilde{\bm{x}}_j)
=
\left|
\left\langle
\phi(\widetilde{\bm{x}}_i)
\middle|
\phi(\widetilde{\bm{x}}_j)
\right\rangle
\right|^{2}.
\label{eq:quantum-kernel}
\end{equation}
The resulting pairwise fidelities form the training and evaluation Gram
matrices, which are supplied as precomputed kernels to a classical support
vector classifier.
For both PCA representations, the validation-selected feature map is
\begin{equation}
U_{\phi}^{(d)}
=
\mathrm{ZZFeatureMap}
\left(d,r=1,\mathcal{E}=\mathrm{full}\right),
\label{eq:qsvm-zzfeaturemap}
\end{equation}
where \(r\) denotes the number of repetitions and \(\mathcal{E}\) specifies
the entanglement pattern. Kernel elements are evaluated exactly using the
Qiskit Machine Learning \texttt{FidelityStatevectorKernel}, without
finite-shot sampling. The resulting Gram matrices are supplied to an SVM
with \(C=1.0\). The decomposed feature-map depths are 17 and 41 for PCA-4
and PCA-8, respectively. These configurations are selected using validation
data and frozen before the final scalability and test evaluations.
Figure~\ref{fig:qsvm-circuit} illustrates the four-qubit quantum-kernel
construction used in the primary PCA-4 analysis.

Algorithm~\ref{alg:qsvm} summarizes the complete QSVM training and evaluation
procedure.

\begin{algorithm}[!t]
\caption{QSVM training and evaluation workflow.}
\label{alg:qsvm}
\footnotesize
\begin{algorithmic}[1]

\Require Training size $N$, seed $r$, PCA dimension $d$
\Ensure Test predictions and evaluation metrics

\State Construct the balanced training subset
$\mathcal{D}_{N}^{(r)}$.

\State Fit preprocessing and PCA on the training subset;
apply the fitted transformations unchanged to validation and test data.

\State Rescale the retained PCA features for quantum encoding.

\State Prepare the quantum feature state
\[
|\phi(\mathbf{x})\rangle
=
U_{\phi}(\mathbf{x})|0\rangle^{\otimes d}.
\]

\State Compute the fidelity kernel
\[
K_{ij}
=
|\langle\phi(\mathbf{x}_i)
|\phi(\mathbf{x}_j)\rangle|^{2}.
\]

\State Construct the training and validation kernel matrices.

\State Train the support vector classifier using the
precomputed training kernel.

\State Obtain validation decision scores and evaluate
BA, weighted ROC--AUC, and macro-$F_{1}$.

\State Select $\tau_{\mathrm{AMS}}^{\star}$ by maximizing
validation-set AMS.

\State Apply the frozen QSVM and fixed threshold
$\tau_{\mathrm{AMS}}^{\star}$ to the test subset.

\State Report test metrics as mean $\pm$ standard deviation
across independent seeds.

\end{algorithmic}
\end{algorithm}

\subsubsection{Variational quantum classifier}
\label{subsec:vqc}

The VQC provides the trainable quantum model in the present analysis.
Unlike the QSVM, whose feature map defines a fixed kernel, the VQC optimizes
a parameterized quantum circuit to discriminate between the two event classes
~\cite{wu2021vqc,belis2021higgs,cerezo2021vqa}. For an input
\(\bm{x}\), the variational quantum state is written as
\begin{equation}
\left|\psi(\bm{x},\bm{\theta})\right\rangle
=
U_{\mathrm{var}}(\bm{\theta})
U_{\phi}(\bm{x})
\left|0\right\rangle^{\otimes d},
\label{eq:vqc-state}
\end{equation}
where \(U_{\phi}(\bm{x})\) denotes the data-encoding circuit,
\(U_{\mathrm{var}}(\bm{\theta})\) is the trainable variational ansatz, and
\(\bm{\theta}\) denotes its trainable parameters.
Computational-basis outcomes are mapped to the two class labels using a
binary parity interpretation. The probability assigned to class
\(c\in\{0,1\}\) is
\begin{equation}
p_{\bm{\theta}}(c\mid\bm{x})
=
\sum_{\bm{z}\in\mathcal{B}_{c}}
\left|
\left\langle
\bm{z}
\middle|
\psi(\bm{x},\bm{\theta})
\right\rangle
\right|^{2},
\label{eq:vqc-class-probability}
\end{equation}
where
\begin{equation}
\mathcal{B}_{c}
=
\left\{
\bm{z}\in\{0,1\}^{d}
:
\operatorname{parity}(\bm{z})=c
\right\}.
\label{eq:vqc-parity-set}
\end{equation}
The signal score used for event ranking and operating-threshold selection is
\(p_{\bm{\theta}}(1\mid\bm{x})\).

The circuit parameters are optimized by minimizing the binary cross-entropy
(BCE) loss,
\begin{equation}
\begin{aligned}
\mathcal{L}_{\mathrm{BCE}}(\bm{\theta})
={}&
-\frac{1}{N_{\mathrm{train}}}
\sum_{i=1}^{N_{\mathrm{train}}}
\Big[
y_i\ln p_{\bm{\theta}}(1\mid\bm{x}_i)
\\
&\qquad
+
(1-y_i)\ln p_{\bm{\theta}}(0\mid\bm{x}_i)
\Big],
\end{aligned}
\label{eq:vqc-bce-loss}
\end{equation}
where \(y_i\in\{0,1\}\) is the true class label. The optimized parameter
vector is therefore
\begin{equation}
\bm{\theta}^{*}
=
\operatorname*{arg\,min}_{\bm{\theta}}
\mathcal{L}_{\mathrm{BCE}}(\bm{\theta}).
\label{eq:vqc-optimization}
\end{equation}
The BCE objective is recorded throughout optimization to assess the
convergence behaviour of the VQC.

\subsubsection{VQC optimization and frozen configurations}
\label{subsec:vqc-optimization}

The VQC architecture and optimization settings are selected using validation
data only, before the sample-size scalability analysis. To reduce the
computational cost of the architecture search, an initial screening is
performed using a balanced \(N=400\) tuning subset drawn from the
\(N=4000\) training sample. The search considers
\texttt{ZFeatureMap} and \texttt{ZZFeatureMap} encodings,
\texttt{EfficientSU2} and \texttt{RealAmplitudes} ansatz circuits with one
or two repetitions, linear or circular entanglement, and the simultaneous perturbation stochastic approximation (SPSA) and
constrained optimization by linear approximation (COBYLA)
optimizers. 
The leading configurations are subsequently evaluated using the complete
\(N=4000\) balanced training samples for different seeds 
with a maximum optimization budget of 150 objective-function evaluations.
The PCA-4 and PCA-8 architectures are selected separately through this
validation-based procedure and frozen before final test evaluation.
For the primary PCA-4 analysis, the selected VQC consists of a four-qubit
\texttt{ZFeatureMap} with one repetition followed by a two-repetition
\texttt{EfficientSU2} ansatz with circular entanglement. The resulting
circuit contains 24 trainable parameters and has a decomposed depth of 16.
This configuration is used unchanged for all training sizes in the primary
PCA-4 scalability analysis.

\begin{figure}[t]
\centering
\includegraphics[width=0.94\columnwidth]
{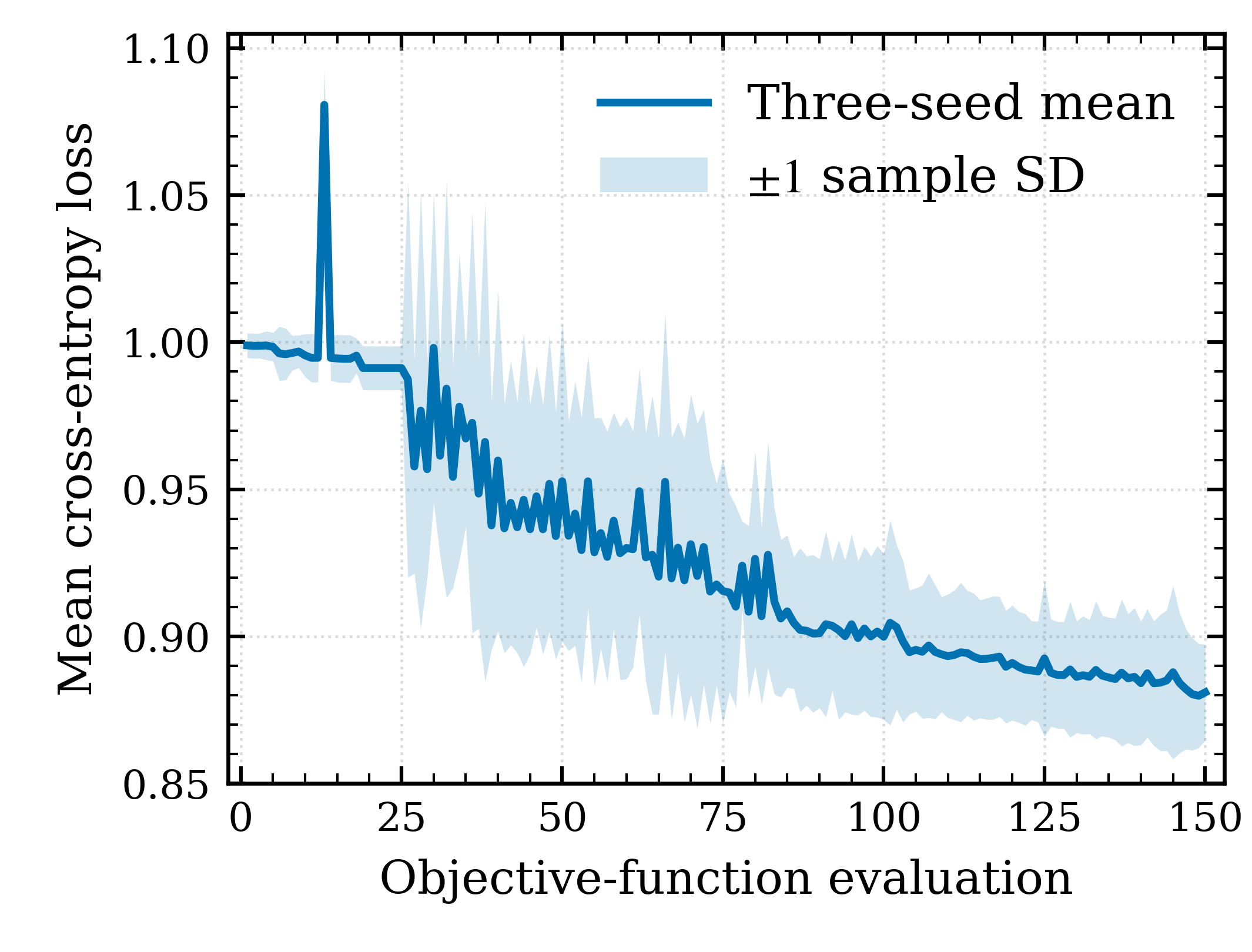}
\caption{
Optimization convergence of the selected PCA-4 VQC at \(N=4000\) for
training seeds \(42\), \(52\), and \(62\). The curves show the binary
cross-entropy objective as a function of the objective-function evaluation.
The selected architecture combines a single-repetition
\texttt{ZFeatureMap} with a two-repetition \texttt{EfficientSU2} ansatz
using circular entanglement and is optimized with COBYLA for a maximum of
150 objective evaluations.
}
\label{fig:vqc-pca4-convergence}
\end{figure}
\begin{figure*}[t]
\centering
\includegraphics[width=0.94\textwidth]
{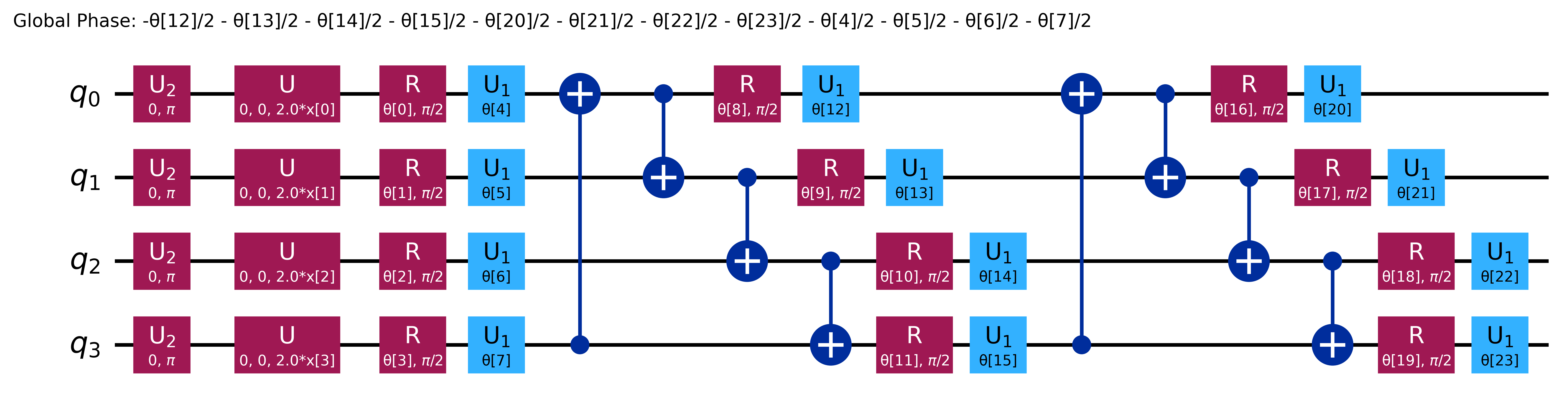}
\caption{
Gate-level decomposition of the four-qubit VQC used in the primary PCA-4
analysis. A single-repetition \texttt{ZFeatureMap} encodes the four rescaled
PCA components, followed by a two-repetition \texttt{EfficientSU2}
variational ansatz with circular entanglement. The variational circuit
contains 24 trainable parameters.
}
\label{fig:vqc-circuit}
\end{figure*}

Figure~\ref{fig:vqc-pca4-convergence} shows the corresponding optimization
trajectories for the three training seeds. The objective decreases across
all three runs, with the mean loss decreasing from approximately \(0.999\)
at initialization to \(0.881\) after optimization.
Figure~\ref{fig:vqc-circuit} shows the decomposed circuit used in the primary
PCA-4 analysis.

For the PCA-8 analysis, the selected VQC consists of an eight-qubit
\texttt{ZZFeatureMap} with one repetition and linear entanglement, followed
by a two-repetition \texttt{RealAmplitudes} ansatz with linear entanglement.
This circuit also contains 24 trainable parameters and has a decomposed depth
of 30.
Both VQC configurations are optimized using COBYLA~\cite{powell1994cobyla}, with a maximum of 150 objective-function
evaluations, an initial trust-region radius of \(0.5\), and a convergence
tolerance of \(10^{-3}\). For each run, the initial parameter vector is
generated deterministically from the corresponding training seed, with
parameter values sampled uniformly from \([-0.1,0.1]\). Class probabilities
are evaluated using the Qiskit reference sampler without finite-shot sampling~\cite{javadiabhari2024qiskit}.
Because the PCA-4 and PCA-8 VQC architectures are selected independently
using validation data and frozen before test evaluation, the secondary
analysis compares the complete representation-specific VQC pipelines rather
than changing the PCA dimension while holding the circuit architecture fixed.
Algorithm~\ref{alg:vqc} summarizes the VQC training and inference procedure.

\begin{algorithm}[!t]
\caption{VQC training and evaluation workflow.}
\label{alg:vqc}
\footnotesize
\begin{algorithmic}[1]

\Require Training size $N$, seed $r$, PCA dimension $d$
\Ensure Test predictions and evaluation metrics

\State Construct the balanced training subset
$\mathcal{D}_{N}^{(r)}$.

\State Fit preprocessing and PCA on the training subset;
apply the fitted transformations unchanged to validation and test data.

\State Rescale the retained PCA features to $[0,\pi]$.

\State Prepare the variational quantum state
\[
|\psi(\mathbf{x},\boldsymbol{\theta})\rangle
=
U_{\mathrm{var}}(\boldsymbol{\theta})
U_{\phi}(\mathbf{x})
|0\rangle^{\otimes d}.
\]

\State Initialize the variational parameters
$\boldsymbol{\theta}$ using the corresponding training seed.

\State Optimize $\boldsymbol{\theta}$ by minimizing binary
cross-entropy loss on the training subset.

\State Obtain validation signal probabilities using the
parity-based measurement rule.

\State Evaluate BA, weighted ROC--AUC, and macro-$F_{1}$ on
the validation subset.

\State Select $\tau_{\mathrm{AMS}}^{\star}$ by maximizing
validation-set AMS.

\State Apply the optimized parameters
$\boldsymbol{\theta}^{*}$ and fixed threshold
$\tau_{\mathrm{AMS}}^{\star}$ to the test subset.

\State Report test metrics as mean $\pm$ standard deviation
across independent seeds.

\end{algorithmic}
\end{algorithm}

\subsection{Evaluation protocol and metrics}
\label{subsec:metrics}

Previous quantum-machine-learning studies of Higgs event classification have
commonly assessed signal--background discrimination using classification
accuracy, receiver operating characteristic (ROC) curves, and the 
corresponding area under the curve (ROC--AUC)
~\cite{mott2017higgs,belis2021higgs,wu2021vqc,wu2021qkernel,
fadol2024cepc,belis2024guided,odagiu2025representations,
maulana2026circuit,aithaddou2026hybrid}.
In the present analysis, classifier performance is evaluated using
complementary class-balanced, threshold-independent, event-weighted, and
physics-oriented measures. The reported metrics are balanced accuracy,
macro-\(F_{1}\), event-weighted ROC--AUC, and approximate median
significance (AMS).
For every classifier, the continuous prediction score
\(f_i=f(\bm{x}_i)\) is defined such that larger values correspond to more
signal-like events. For classifiers with probabilistic outputs,
\(f_i\) is the predicted signal probability. For the RBF--SVM and QSVM,
\(f_i\) is the signed decision-function value. Binary predictions used to
compute balanced accuracy and macro-\(F_{1}\) follow the standard decision
rule of each model: a threshold of \(0.5\) for probabilistic outputs and
zero for signed SVM margins. Event weights are not used in the calculation
of balanced accuracy or macro-\(F_{1}\).
ROC--AUC is evaluated directly from the continuous classifier scores and is
therefore independent of a fixed classification threshold. In addition to
the standard unweighted ROC--AUC, an event-weighted ROC--AUC is computed
using the event weights supplied with the dataset, thereby incorporating
the relative statistical contribution of each simulated event. The AMS is
evaluated separately at an operating threshold selected exclusively from
validation data and is not itself an asymptotic likelihood-based discovery
significance~\cite{cowan2011asymptotic}.

\paragraph{Balanced accuracy and macro-\texorpdfstring{\(F_{1}\)}{F1}.}
Balanced accuracy is used because the validation and test samples retain the
natural signal--background imbalance of the dataset. It assigns equal
importance to the recall of the two classes and is defined as
\cite{brodersen2010balanced}
\begin{equation}
\mathrm{BA}
=
\frac{1}{2}
\left(
\frac{\mathrm{TP}}{\mathrm{TP}+\mathrm{FN}}
+
\frac{\mathrm{TN}}{\mathrm{TN}+\mathrm{FP}}
\right),
\label{eq:balanced-accuracy}
\end{equation}
where \(\mathrm{TP}\) and \(\mathrm{FN}\) denote correctly and incorrectly
classified signal events, while \(\mathrm{TN}\) and \(\mathrm{FP}\) denote
correctly and incorrectly classified background events.

For each class \(c\in\{0,1\}\), precision and recall are
\begin{equation}
P_c
=
\frac{\mathrm{TP}_c}
{\mathrm{TP}_c+\mathrm{FP}_c},
\qquad
R_c
=
\frac{\mathrm{TP}_c}
{\mathrm{TP}_c+\mathrm{FN}_c},
\label{eq:precision-recall}
\end{equation}
where the confusion-matrix entries are defined in a one-versus-rest manner
~\cite{sokolova2009performance}.
The class-specific \(F_{1}\) score is
\begin{equation}
F_{1,c}
=
\frac{2P_cR_c}{P_c+R_c},
\label{eq:class-f1}
\end{equation}
and the macro-averaged score is
\begin{equation}
F_{1,\mathrm{macro}}
=
\frac{1}{2}
\left(
F_{1,0}+F_{1,1}
\right).
\label{eq:macro-f1}
\end{equation}
Macro-\(F_{1}\) therefore gives equal importance to signal and background
classification while jointly accounting for precision and recall.

\paragraph{Event-weighted ROC--AUC}

ROC--AUC quantifies the ability of a classifier to rank signal events above
background events over all possible score thresholds. Because the ATLAS
Higgs Challenge events carry event weights, both standard and event-weighted
ROC--AUC are evaluated~\cite{fawcett2006roc}. The weighted empirical ROC--AUC
is defined as
\begin{equation}
\mathrm{AUC}_{w}
=
\frac{
\displaystyle
\sum_{i:y_i=1}
\sum_{j:y_j=0}
w_iw_j
\left[
\mathbf{1}(f_i>f_j)
+
\frac{1}{2}\mathbf{1}(f_i=f_j)
\right]
}{
\displaystyle
\left(
\sum_{i:y_i=1}w_i
\right)
\left(
\sum_{j:y_j=0}w_j
\right)
},
\label{eq:weighted-roc-auc}
\end{equation}
where \(w_i\) denotes the event weight and
\(\mathbf{1}(\cdot)\) is the indicator function. The standard unweighted
ROC--AUC is recovered by assigning unit weight to every event.
A value of \(0.5\) corresponds to random ranking, whereas values approaching
unity indicate increasingly effective signal--background discrimination.
ROC--AUC measures event-level ranking performance, while its weighted form
accounts for the relative statistical contribution of the simulated events.
Neither metric requires a fixed binary operating threshold.

\paragraph{Approximate median significance.}

AMS is used to assess the physics-oriented quality of a signal-enriched event
selection and follows the Higgs Machine Learning Challenge convention
~\cite{adambourdarios2015higgs}. Because the matched validation and test
samples contain only 800 events, their event weights are expanded to account
for the class-wise sampling fractions. For a partition
\(D\in\{\mathrm{val},\mathrm{test}\}\), the expansion factor for class
\(c\in\{0,1\}\) is
\begin{equation}
\alpha_{c}^{(D)}
=
\frac{N_{c}^{(D)}}{n_{c}^{(D)}},
\label{eq:ams-expansion-factor}
\end{equation}
where \(N_{c}^{(D)}\) is the number of class-\(c\) events in the complete
permanent partition and \(n_{c}^{(D)}\) is the corresponding number in the
matched subset. The expanded event weight is therefore
\begin{equation}
w_{i,\mathrm{exp}}^{(D)}
=
\alpha_{y_i}^{(D)} w_i.
\label{eq:expanded-event-weight}
\end{equation}

For a score threshold \(\tau\), the corresponding weighted signal and
background yields are
\begin{align}
s_D(\tau)
&=
\sum_{i\in D_{\mathrm{match}}}
w_{i,\mathrm{exp}}^{(D)}
\mathbf{1}(y_i=1)
\mathbf{1}(f_i\geq\tau),
\label{eq:ams-signal-yield}
\\
b_D(\tau)
&=
\sum_{i\in D_{\mathrm{match}}}
w_{i,\mathrm{exp}}^{(D)}
\mathbf{1}(y_i=0)
\mathbf{1}(f_i\geq\tau).
\label{eq:ams-background-yield}
\end{align}
Here, \(b_D(\tau)\) denotes the weighted background yield passing the
signal-selection threshold.
The AMS is then calculated as
\begin{equation}
\begin{aligned}
\mathrm{AMS}_{D}(\tau)
&=
\Bigg\{
2\Bigg[
\bigl(s_D(\tau)+b_D(\tau)+b_r\bigr)
\\
&\qquad \times
\ln\!\left(
1+
\frac{s_D(\tau)}
     {b_D(\tau)+b_r}
\right)
-s_D(\tau)
\Bigg]
\Bigg\}^{1/2}.
\end{aligned}
\label{eq:ams}
\end{equation}
with the regularization term fixed to \(b_r=10\), following the HiggsML
convention. The validation-selected threshold follows the two-stage
training--validation procedure used for AMS optimization~\cite{kotlowski2015ams}.
A separate operating threshold is determined for every classifier,
representation, training size, and training seed by maximizing the AMS on
the corresponding validation sample,
\begin{equation}
\tau_{\mathrm{AMS}}^{\star}
=
\underset{\tau}{\operatorname{arg\,max}}\,
\mathrm{AMS}_{\mathrm{val}}(\tau),
\label{eq:ams-validation-threshold}
\end{equation}
and is subsequently transferred unchanged to the corresponding test sample.
The resulting test significance is therefore
\begin{equation}
\mathrm{AMS}_{\mathrm{test}}^{\star}
=
\mathrm{AMS}_{\mathrm{test}}
\!\left(\tau_{\mathrm{AMS}}^{\star}\right).
\label{eq:transferred-test-ams}
\end{equation}
No threshold optimization is performed using the test events.

\paragraph{Aggregation across training seeds.}

Each metric is calculated independently for training seeds
\(r\in\{42,52,62\}\). For a metric \(M\) evaluated over \(R=3\) seeds,
the reported mean and sample standard deviation are
\begin{align}
\overline{M}
&=
\frac{1}{R}
\sum_{r=1}^{R}M_r,
\label{eq:metric-mean}
\\
s_M
&=
\sqrt{
\frac{1}{R-1}
\sum_{r=1}^{R}
\left(
M_r-\overline{M}
\right)^2
}.
\label{eq:metric-standard-deviation}
\end{align}
Results are reported as \(\overline{M}\pm s_M\). For the secondary PCA-4
versus PCA-8 analysis, paired metric differences are first calculated within
each training seed and subsequently aggregated across seeds. Because only
three seeds are considered, the reported standard deviations are interpreted
as descriptive measures of seed-dependent variability rather than as
inferential uncertainty estimates.
Balanced accuracy and macro-\(F_{1}\) characterize performance under a fixed
binary decision rule, whereas ROC--AUC assesses global score ranking and
event-weighted ROC--AUC incorporates the supplied event weights. AMS provides
a complementary physics-oriented measure based on the weighted signal and
background yields at a validation-selected operating point. The expanded
matched-test AMS is used for controlled relative comparison and is not
interpreted as an exact full-partition AMS, an official HiggsML leaderboard
score, or a standalone experimental discovery significance.

\section{Results and discussion}
\label{sec:results-discussions}

The results are organized around two central questions: how classical and quantum classifier performance scales with the available 
training statistics, and how this comparison depends on the dimensionality of the input representation. All representation and model-selection 
decisions are based exclusively on validation data, while the independent test partition is reserved for final evaluation.
We first report the validation-based representation screening that motivates the use of PCA-4 as the reference representation 
for the primary scalability study. Under this common four-dimensional input space, logistic regression, RBF--SVM, random forest, XGBoost, QSVM, and VQC 
are evaluated on balanced training subsets \ref{eq:nested-training}.
The final configurations are then compared in detail, followed by a paired PCA-4 versus PCA-8 analysis using matched training samples and common evaluation events to isolate the effect of representation dimension. Finally, the classification results are complemented by the approximate median significance (AMS), with operating thresholds determined on validation data and applied unchanged to the independent test sample.
Unless stated otherwise, results are reported as the mean ($\pm$) sample standard deviation over seeds (r). The validation and test partitions remain fixed across these repetitions, ensuring that all reported test results correspond to genuinely out-of-sample performance.

\subsection{Representation screening and selection}
\label{subsec:representation-results}

\begin{figure*}[!t]
\centering
\includegraphics[width=0.94\textwidth]{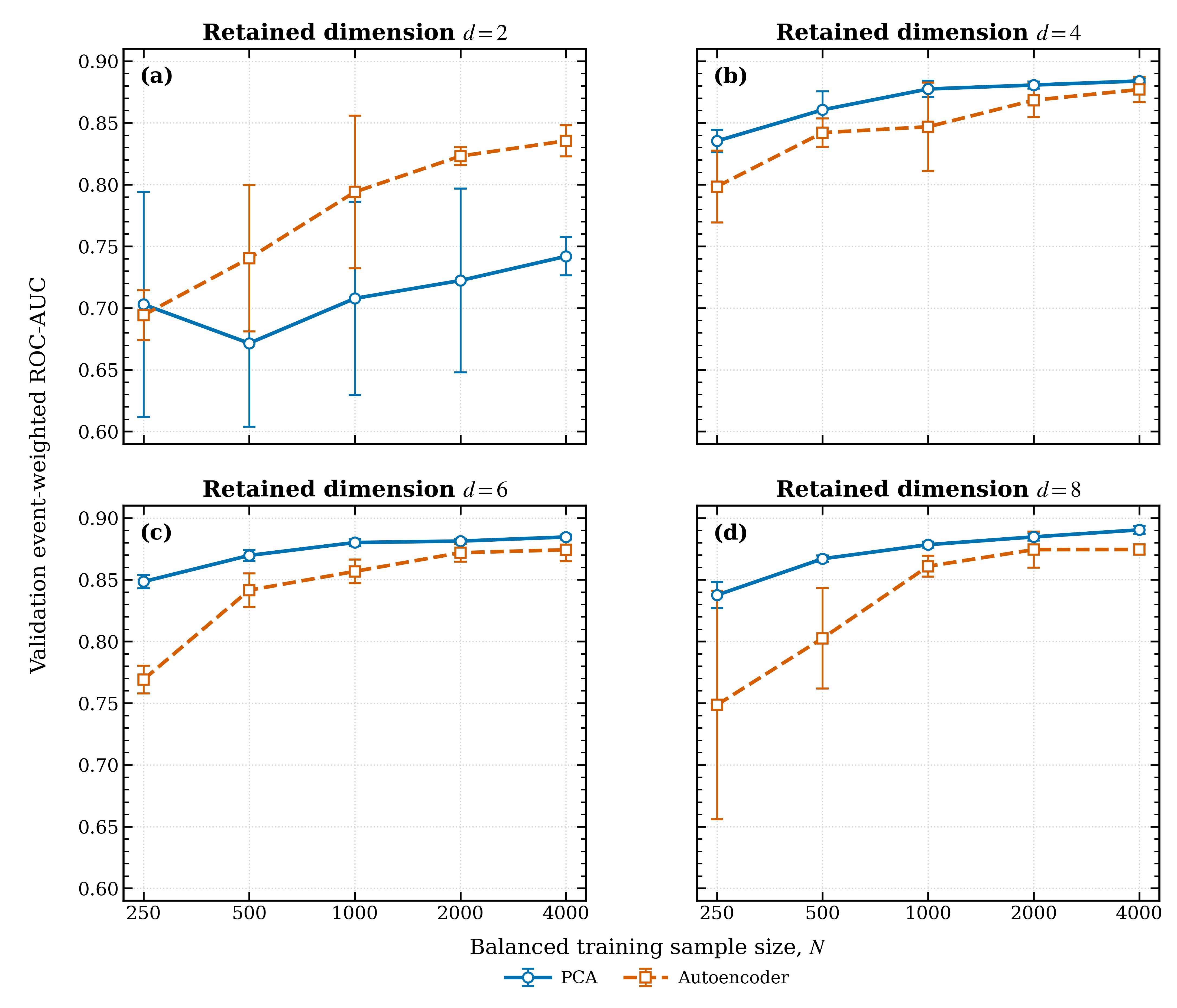}
\caption{
Validation event-weighted ROC--AUC of the RBF--SVM for PCA and autoencoder
representations as a function of the balanced training-set size. Markers show
the mean over seeds (r) and error bars denote the
sample standard deviation. Averaging is performed independently at each
training size.
}
\label{fig:pca-ae-screening}
\end{figure*}

PCA and AE representations with $d\in\{2,4,6,8\}$
were evaluated using validation performance, information retention or reconstruction quality, cross-seed stability, and 
quantum-resource feasibility. For each training size, results were averaged independently over seeds (r); values from 
different training sizes were not pooled.
Figure~\ref{fig:pca-ae-screening} compares the validation event-weighted ROC--AUC of the RBF--SVM across the investigated representations. 
The RBF--SVM was used as a representative nonlinear downstream classifier because it could be evaluated consistently over the complete representation sweep. 
The final representation choice, however, was based on the full screening criteria rather than on this classifier alone. Table~\ref{tab:pca-ae-other-classifiers} compares the downstream classification performance of PCA and AE representations for three additional classical 
classifiers at $N=4000$.
The preferred reduction method depends on both the classifier and the retained dimension.

\begin{table}[t]
\centering
\caption{Validation event-weighted ROC--AUC for PCA and autoencoder
(AE) representations at $N=4000$. Values are mean $\pm$ standard
deviation over seeds. Boldface indicates the higher
mean within each classifier--dimension pair.}
\label{tab:pca-ae-other-classifiers}

\footnotesize
\setlength{\tabcolsep}{3.5pt}
\renewcommand{\arraystretch}{1.05}

\begin{tabular}{lccc}
\toprule
Classifier & $d$ & PCA & AE \\
\midrule

LR
& 2 & $0.7025\pm0.0094$ & {\boldmath$0.7355\pm0.0254$} \\
& 4 & $0.8189\pm0.0036$ & {\boldmath$0.8264\pm0.0147$} \\
& 6 & {\boldmath$0.8201\pm0.0038$} & $0.7965\pm0.0550$ \\
& 8 & {\boldmath$0.8308\pm0.0025$} & $0.7813\pm0.0337$ \\

\midrule

RF
& 2 & $0.6817\pm0.0182$ & {\boldmath$0.8178\pm0.0067$} \\
& 4 & {\boldmath$0.8690\pm0.0017$} & $0.8631\pm0.0047$ \\
& 6 & {\boldmath$0.8781\pm0.0031$} & $0.8639\pm0.0168$ \\
& 8 & {\boldmath$0.8896\pm0.0033$} & $0.8678\pm0.0128$ \\

\midrule

XGB
& 2 & $0.7195\pm0.0203$ & {\boldmath$0.8416\pm0.0086$} \\
& 4 & {\boldmath$0.8845\pm0.0016$} & $0.8730\pm0.0112$ \\
& 6 & {\boldmath$0.8846\pm0.0021$} & $0.8654\pm0.0169$ \\
& 8 & {\boldmath$0.8912\pm0.0020$} & $0.8701\pm0.0123$ \\

\bottomrule
\end{tabular}
\end{table}
At $d=2$, the autoencoder achieves a higher mean event-weighted ROC--AUC for all three classifiers.
For random forest and XGBoost, PCA yields the higher mean at $d=4,6,8$. Logistic regression shows a mixed response: the autoencoder performs better at $d=2,4$, whereas PCA performs 
better at $d=6,8$. At $d=4,6,8$, PCA also has a smaller sample standard deviation across training seeds for each of the three classifiers. 
These comparisons support a classifier- and dimension-dependent interpretation of representation quality.
They provide supporting evidence for using PCA as a common representation, while the choice of four components additionally reflects the quantum-resource constraint.
The screening shows that (d=2) generally removes too much discriminative information,
whereas representations with $(d\geq4)$ recover substantially stronger downstream performance. 
Across the low-dimensional regime, PCA provided the most consistent balance between classification performance, cross-seed stability, and 
computational simplicity, 
while the AE representations showed no systematic validation advantage 
sufficient to justify their additional training dependence.
PCA-4 was therefore fixed for the primary scalability analysis. It retains approximately half of the variance of the preprocessed feature space 
while restricting the quantum models to four qubits and providing an identical input representation for all six classifiers. This choice should be interpreted as a common, 
computationally tractable reference representation rather than as the universally optimal dimensionality for every classifier. 
PCA-4 versus PCA-8 was retained for the secondary representation-sensitivity analysis, allowing the effect of increased input dimensionality to be assessed separately from the primary scalability study.

\subsection{Primary PCA-4 sample-size scalability}
\label{subsec:pca4-scalability-results}

\begin{figure*}[t]
    \centering
    \includegraphics[width=0.94\textwidth]
    {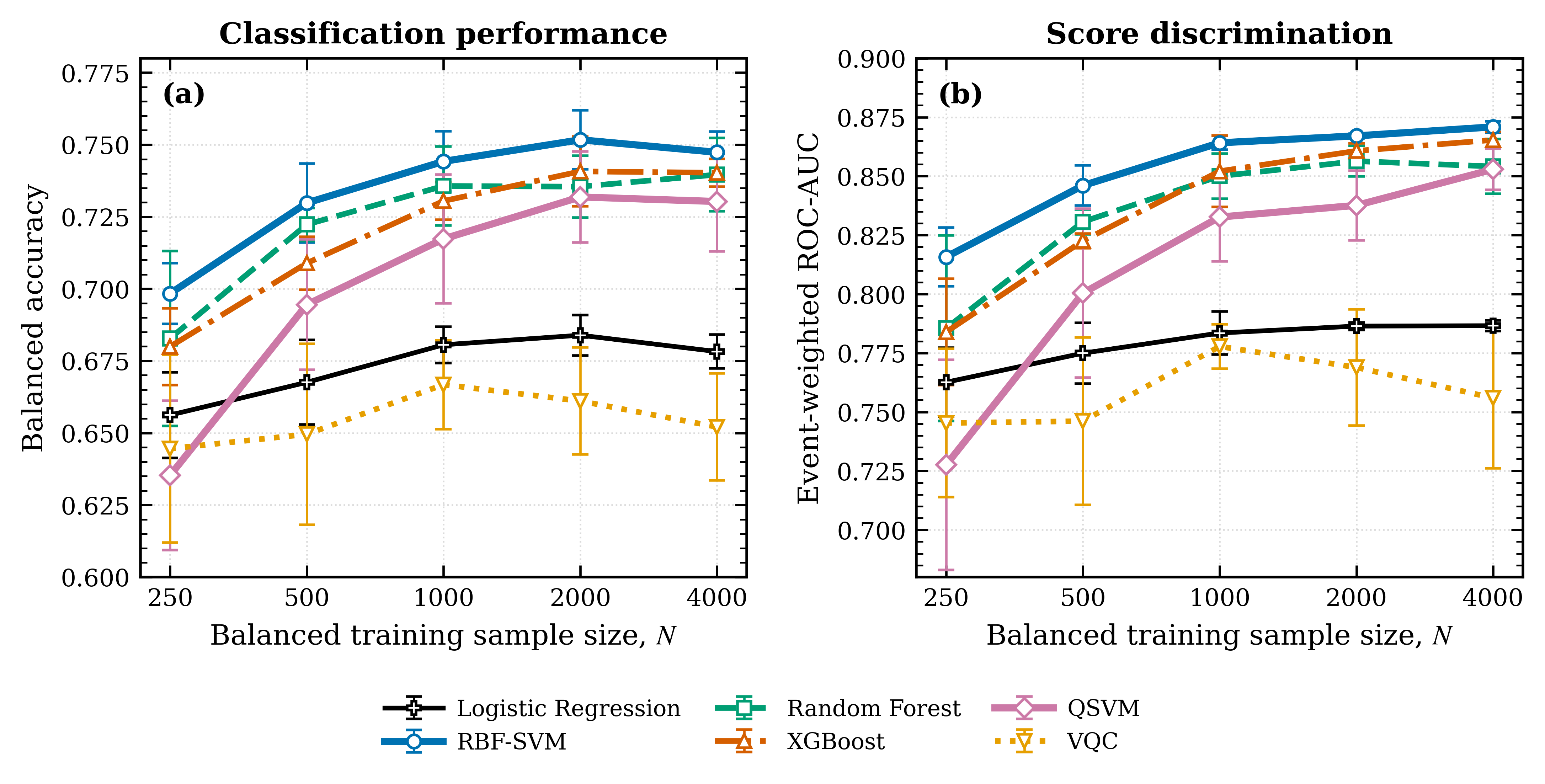}
    \caption{
    Sample-size scalability of the classical and quantum classifiers under
    the primary PCA-4 representation. Panel (a) shows balanced accuracy and
    panel (b) the event-weighted ROC--AUC on the common independent test
    sample. Points denote the mean over training seeds 42, 52, and 62, and
    error bars show the sample standard deviation. Model configurations
    remain fixed across all training sizes.
    }
    \label{fig:pca4-scalability}
\end{figure*}

The primary analysis examines the sample-size dependence of all six
classifiers under the common PCA-4 representation.
Figure~\ref{fig:pca4-scalability} compares balanced accuracy and event-weighted ROC-AUC as functions of the balanced training-set size $N$ under
primary PCA-4 protocol.
Among the classical classifiers, the RBF--SVM achieves the highest observed
mean balanced accuracy and event-weighted ROC--AUC at all investigated values
of $N$. Random forest and XGBoost improve overall with increasing training
size and approach the leading classical benchmark at larger values of $N$,
although their curves are not strictly monotonic in both metrics. Logistic
regression improves modestly between smallest and largest training samples, but it still remains below these nonlinear
classifiers.
Among the quantum classifiers, the QSVM exhibits a substantial improvement
between the smallest and largest training subsets. 
The differing QSVM trends in the two panels are not contradictory in Figure~\ref{fig:pca4-scalability}.
Balanced accuracy evaluates binary predictions at the prescribed decision
threshold, whereas event-weighted ROC--AUC evaluates weighted
signal--background score ordering across thresholds
~\cite{fawcett2006roc}. At $N=250$, the mean $\pm$ sample-standard-deviation intervals for QSVM and VQC overlap. The observed difference in mean performance should therefore 
be interpreted cautiously in light of variability across training seeds. These descriptive error bars alone do not establish a statistically significant performance advantage for 
either model at this training 
size. 
However, the mean event-weighted
ROC--AUC and mean balanced accuracy of QSVM increase with the larger training sizes. 
These observations indicate that the selected quantum-kernel pipeline
benefits more consistently from additional training data than the selected
variational pipeline. 

A relevant methodological distinction is that the QSVM uses the quantum
circuit to define a fidelity kernel and learns the decision function through
a classical support vector machine
~\cite{havlicek2019supervised,schuld2019feature}. For a fixed
positive-semidefinite kernel, this training problem is convex
~\cite{cortes1995} and does not require the nonconvex circuit-parameter
optimization which is used by VQC. The observed QSVM improvement is therefore
consistent with additional labelled events supporting a more effective
separating function in the encoded feature space. However, convexity
guarantees neither low test error nor an advantage over a classical kernel.
Moreover, the QSVM and VQC use different encodings, so their performance
difference cannot be attributed solely to the optimization method.

The comparison between the RBF--SVM and QSVM further illustrates the
remaining performance gap between the leading classical and quantum-kernel
classifiers. At $N=4000$, the RBF--SVM exceeds the QSVM by $0.0171$ in mean
balanced accuracy and $0.0180$ in mean event-weighted ROC--AUC. Although
these gaps are smaller than at $N=250$, the five investigated training sizes
do not establish eventual convergence or a future crossover. The residual
gap to the RBF--SVM is consistent with the task dependence of quantum-kernel
performance. Generalization depends on the relationship between the
kernel-induced function class and the classification target, rather than on
feature-space dimension alone
~\cite{fig6_kubler2021,huang2021,Thanasilp2024Concentration}. Encoding and
input scaling are consequently relevant modelling choices
~\cite{shaydulin2022}. Related LHC studies have evaluated quantum-kernel
classifiers on substantially larger simulated samples and reported
performance comparable to classical counterparts; however, differences in
datasets and feature maps prevent a direct quantitative comparison
~\cite{wu2021qkernel}.

In contrast to the QSVM, both VQC mean metrics reach their highest plotted
values at $N=1000$ and subsequently decrease. Exact simulation excludes
hardware noise and finite-shot sampling as explanations. Plausible
explanations include limited variational optimization, restrictions 
imposed by the encoding, and changes in the training-fitted
representation
~\cite{anschuetz2022,Schuld2021Encoding,Cerezo2021Cost}. The figure does not
distinguish these possibilities or establish overfitting, a barren plateau,
or a general deterioration caused by additional training data.
All comparisons concern the specified training pipelines: preprocessing and
PCA are refitted for each training subset, while the model configurations are
frozen. The error bars describe sample standard deviations across three
training seeds on a common fixed test set. They are not confidence intervals
for paired differences, and their overlap does not establish statistical
equivalence~\cite{cumming2007}. Accordingly, the figure supports improved
observed QSVM discrimination and a reduced endpoint gap to the leading
classical benchmark, but does not demonstrate predictive quantum advantage
in the investigated regime. 
\begin{table*}[t]
    \centering
    \caption{
    Performance of the six classifiers at \(N=4000\) under the primary
    PCA-4 protocol. Values are the mean \(\pm\) sample standard deviation
    over training seeds 42, 52, and 62, evaluated on the common fixed
    800-event test sample. AMS operating thresholds were determined from
    validation data and transferred unchanged to the test sample. Boldface
    denotes the highest mean value for each metric.
    }
    \label{tab:pca4-n4000-performance}
    \begin{tabular}{lcccc}
        \hline
        Model &
        Balanced accuracy &
        Weighted ROC--AUC &
        Macro-F1 &
        Expanded AMS \\
        \hline

        RBF--SVM
        & \(\mathbf{0.7474\pm0.0072}\)
        & \(\mathbf{0.8709\pm0.0024}\)
        & \(\mathbf{0.7263\pm0.0111}\)
        & \(0.8456\pm0.1170\) \\

        XGBoost
        & \(0.7403\pm0.0048\)
        & \(0.8654\pm0.0034\)
        & \(0.7228\pm0.0065\)
        & \(\mathbf{1.1053\pm0.1946}\) \\

        Random forest
        & \(0.7396\pm0.0127\)
        & \(0.8541\pm0.0116\)
        & \(0.7249\pm0.0139\)
        & \(0.8692\pm0.2704\) \\

        QSVM
        & \(0.7303\pm0.0173\)
        & \(0.8529\pm0.0088\)
        & \(0.7092\pm0.0192\)
        & \(0.8921\pm0.0120\) \\

        Logistic regression
        & \(0.6782\pm0.0059\)
        & \(0.7866\pm0.0022\)
        & \(0.6661\pm0.0064\)
        & \(0.5840\pm0.0052\) \\

        VQC
        & \(0.6521\pm0.0186\)
        & \(0.7559\pm0.0299\)
        & \(0.6404\pm0.0086\)
        & \(0.5720\pm0.0407\) \\
        \hline
    \end{tabular}
\end{table*}
The AMS results provide a complementary physics-oriented perspective. The classifier ranking depends on the evaluation 
objective. At \(N=4000\), RBF–SVM achieves the highest mean balanced accuracy, 
weighted ROC–AUC and macro-F1, whereas XGBoost attains the highest mean expanded AMS as shown in Table~\ref{tab:pca4-n4000-performance}. This difference is consistent with the 
distinction between discrimination integrated over score thresholds and sensitivity evaluated from the weighted signal and background yields at a selected operating point. 
The QSVM achieves the second-highest mean expanded AMS, exceeding the corresponding mean values of random forest and RBF–SVM. Thus, its 
lower global discrimination metrics do not preclude competitive performance for the AMS selection objective. Its comparatively small AMS dispersion indicates limited 
variation across the three evaluated seeds, although this observation does not establish robustness to independent evaluation samples or systematic variations. 
Logistic regression and VQC yield lower mean AMS values under the same protocol.
All AMS thresholds were selected using validation data and subsequently applied unchanged to the common test sample. 
The reported uncertainties represent sample standard deviations across three training seeds; consequently, the ordering of the means 
should be interpreted descriptively, without claims of statistical superiority or quantum advantage. Expanded AMS provides a comparative 
sensitivity measure under the adopted yield normalization, rather than an observed experimental discovery significance.

\subsection{Secondary PCA-4 versus PCA-8 representation analysis}
\label{subsec:representation-ablation-results}

\begin{figure*}[t]
    \centering
    \includegraphics[width=0.94\textwidth]{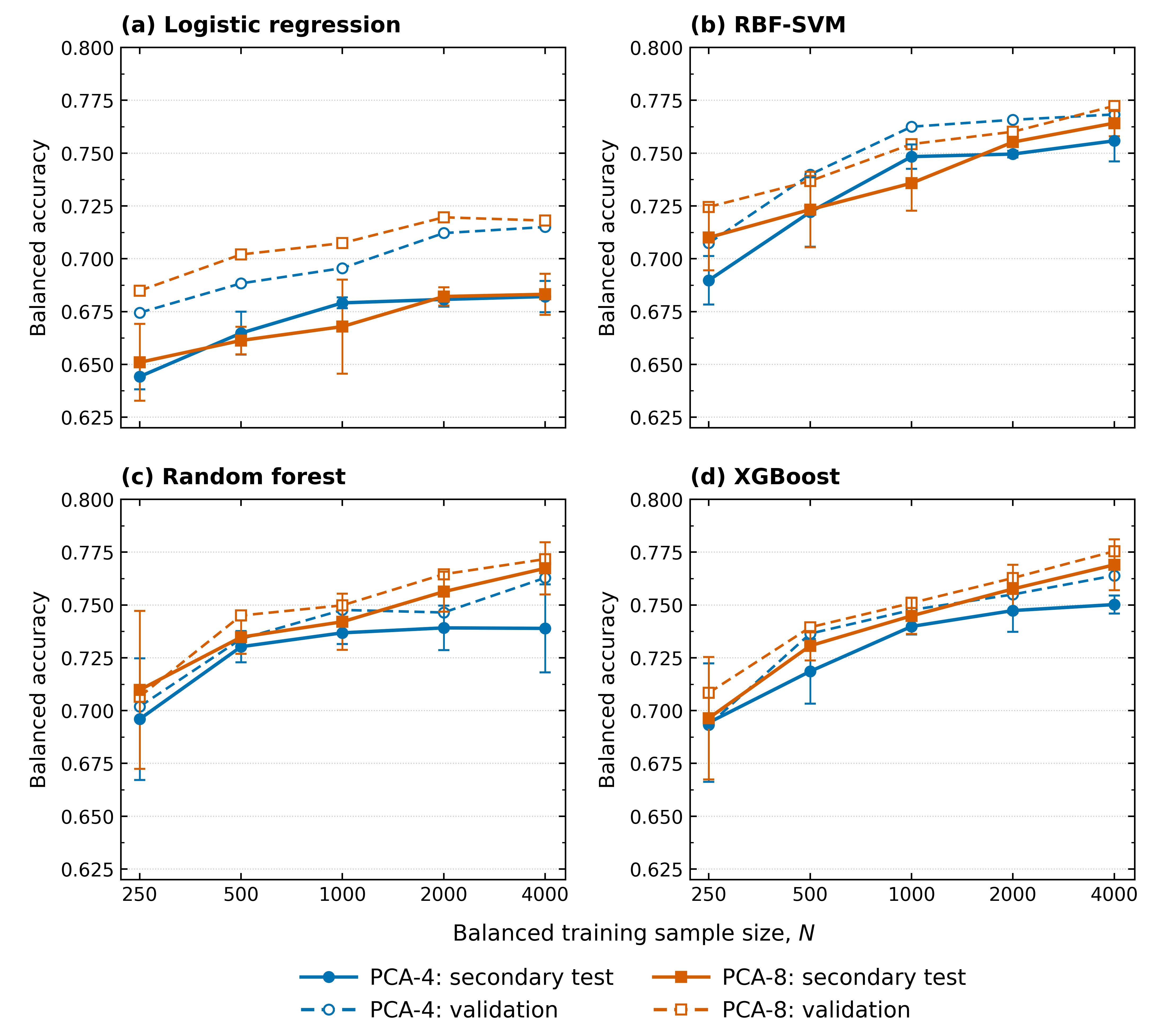}

    \caption{
    Sensitivity of the classical classifiers to the PCA representation
    dimension. Balanced accuracy is shown for PCA-4 and PCA-8 as a
    function of the balanced training-set size. Solid curves denote the
    independent secondary test sample, while dashed curves show the
    corresponding validation results. Test error bars represent the
    sample standard deviation across seeds. PCA-4 and PCA-8 are evaluated
    on matched event identities for every model, training size, and seed.
    }
    \label{fig:pca4-pca8-classical}
\end{figure*}

The secondary analysis examines whether the conclusions of the primary
PCA-4 study are sensitive to an increase in representation dimension. Following
the validation analysis, a separate stratified 800-event subset of the
permanent test partition was prospectively fixed for this comparison
(seed 2028), excluding the events used in the primary PCA-4 test evaluation.
For every model, training size, and seed, the PCA-4 and PCA-8 configurations
were evaluated on identical event identities using the previously frozen
training procedure and model settings. No hyperparameters, classification
thresholds, or representation choices were modified using this secondary test
sample. Paired differences are defined as
\begin{equation}
\Delta M =
M_{\mathrm{PCA\text{-}8}}
-
M_{\mathrm{PCA\text{-}4}},
\label{eq:pca4-pca8-delta}
\end{equation}
so that $\Delta M>0$ favours PCA-8.
Increasing the representation from four to eight components substantially
increases the retained PCA variance. The corresponding classification response,
however, is strongly architecture dependent.
Figure~\ref{fig:pca4-pca8-classical} shows the balanced-accuracy behaviour of
the four classical classifiers on both validation data and the independent
secondary test sample. 
Random forest and XGBoost exhibit the most consistent benefits from PCA-8. At \(N=4000\), 
their balanced accuracies and their weighted ROC–AUC values also increase. Both classifiers favour PCA-8 in mean balanced accuracy at 
all five training sizes, with complete agreement between the validation and secondary-test directions. These observations are 
consistent with the additional components providing useful structure that the tree ensembles can exploit through their feature-dependent partitions. 
They do not imply that every additional component is informative or that performance would continue improving at larger dimensions.
The response is more modest for RBF–SVM, which gains 0.0084 in balanced accuracy and 0.0073 in weighted ROC–AUC at \(N=4000\). 
PCA-8 gives higher mean balanced accuracy at four of five training sizes, although validation and test agree in direction at only three as shown in Table~\ref{tab:pca4-pca8-paired-effects}. 
Logistic regression exhibits almost no change in balanced accuracy at the largest training size, but its weighted ROC–AUC increases. 
Its representation dependence should therefore be described as metric dependent. Improved event-weighted ranking does not necessarily produce a comparable improvement in 
balanced accuracy at the classification threshold. 

\begin{figure*}[t]
    \centering
    \includegraphics[
        width=0.94\textwidth]{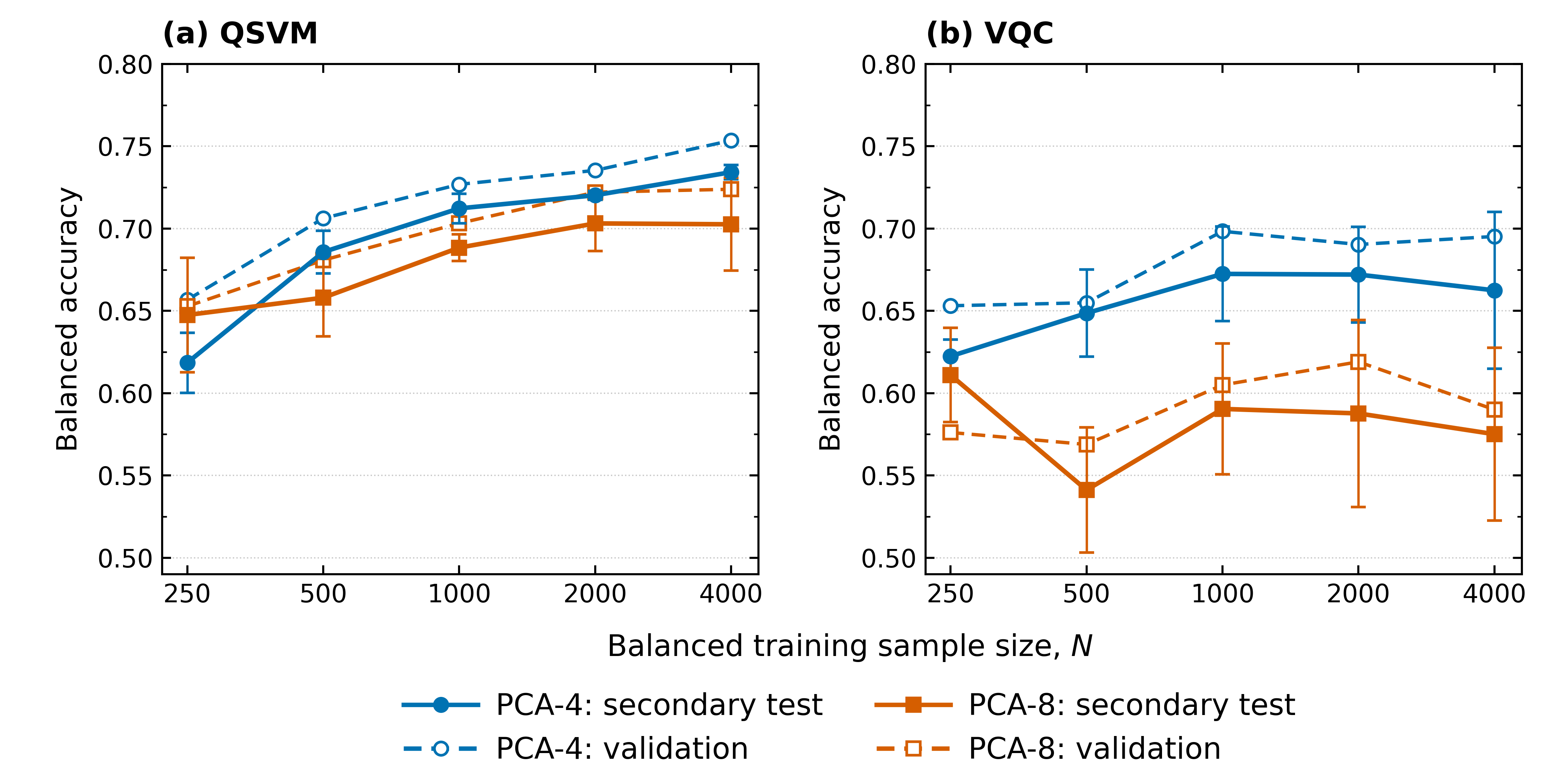}
  \caption{
    Sensitivity of the quantum classifiers to the PCA representation dimension.
    Panels (a) and (b) show the balanced accuracy of the QSVM and VQC,
    respectively, for PCA-4 and PCA-8. Solid curves correspond to the independent
    secondary test sample and dashed curves to validation data. Error bars on the
    test results denote the sample standard deviation over different seeds. The two representations are evaluated using matched training
    subsets and identical evaluation events.
  }
  \label{fig:pca4-pca8-quantum}
\end{figure*}

A qualitatively different dependence is observed for the quantum classifiers,
as shown in Figure~\ref{fig:pca4-pca8-quantum}. The QSVM favours PCA-4 at four 
of the five secondary-test training sizes; only at $N=250$ does PCA-8 give 
a higher mean balanced accuracy. At $N=4000$, the balanced accuracy and event weighted ROC-AUC decreases 
corresponding PCA-8, as compared to PCA-4. The QSVM favours PCA-4 at four of the five secondary-test training sizes; only at \(N=250\) does PCA-8 yield a higher mean balanced accuracy. 
At \(N=4000\), both the balanced accuracy and event-weighted ROC-AUC decrease for PCA-8 compared with PCA-4.
Increasing the PCA dimension simultaneously increases the number of qubits and changes the fidelity kernel induced by the encoding. A possible explanation is 
that the resulting kernel geometry is less well matched to the classification task under the adopted scaling and regularization. 
This interpretation is motivated by studies demonstrating the importance of quantum-kernel bandwidth and task alignment, but remains a 
hypothesis for the present dataset. Because the kernel is evaluated exactly and the subsequent SVM optimization is convex, difficulties in 
optimizing a variational circuit do not explain this QSVM result~\cite{shaydulin2022,Canatar2023Bandwidth}.
VQC exhibits the largest observed deterioration, with differences of \(-0.0874\) in balanced accuracy and \(-0.0876\) in weighted ROC–AUC. 
PCA-4 performs better in mean balanced accuracy at every training size, and the direction agrees between validation 
and test in all five comparisons as illustrated in Table~\ref{tab:pca4-pca8-paired-effects}. 
Since weighted ROC–AUC also decreases, the deterioration cannot be explained solely by the classification threshold. 
Possible contributors include the suitability of the eight-qubit encoding and ansatz for the classification task, as well as degree of convergence 
achieved within the fixed optimization budget. 
Encoding and trainability studies motivate these possibilities, but additional diagnostics are required to distinguish them; the present results alone do not establish 
barren plateaus or overfitting~\cite{Schuld2021Encoding,cerezo2021vqa}. 
The win counts and validation–test agreement summarize directional consistency across the investigated training sizes. They are not independent 
significance tests, because the training subsets are nested and the evaluation events are shared. The results therefore support PCA-8 for the evaluated tree 
ensembles and PCA-4 for the evaluated quantum configurations within this experimental regime. They do not establish a universally optimal representation dimension or an 
intrinsic disadvantage of using more qubits.

The magnitude and direction of the representation effect at the largest
training size are summarized in
Table~\ref{tab:pca4-pca8-paired-effects}. In addition to balanced accuracy,
the weighted ROC--AUC differences confirm the same broad pattern: PCA-8
benefits the tree-based classifiers, produces only small changes for the
RBF--SVM and logistic regression, and reduces the performance of both quantum
models.
\begin{table*}[t]
\centering
\caption{
Paired PCA-8 minus PCA-4 performance differences on the independent secondary
test sample. Differences are defined as
$\Delta M=M_{\mathrm{PCA\text{-}8}}-M_{\mathrm{PCA\text{-}4}}$, where $M$
denotes balanced accuracy (BA) or event-weighted ROC--AUC; positive values
favour PCA-8. Numerical $\Delta M$ values are reported at $N=4000$.
``PCA-8 BA wins'' counts the five training sizes for which PCA-8 has the
higher mean BA, while ``direction agreement'' counts the cases in which the
validation and test differences have the same sign.
}
\label{tab:pca4-pca8-paired-effects}
\begin{tabular}{lcccc}
\toprule
\textbf{Model} &
\shortstack{\textbf{$\Delta$ balanced}\\\textbf{accuracy}} &
\shortstack{\textbf{$\Delta$ weighted}\\\textbf{ROC--AUC}} &
\shortstack{\textbf{PCA-8 BA}\\\textbf{wins}} &
\shortstack{\textbf{Validation/test}\\\textbf{agreement}} \\
\midrule
Logistic regression & $+0.0011$ & $+0.0207$ & $3/5$ & $3/5$ \\
RBF--SVM            & $+0.0084$ & $+0.0073$ & $4/5$ & $3/5$ \\
Random forest       & $+0.0284$ & $+0.0237$ & $5/5$ & $5/5$ \\
XGBoost             & $+0.0188$ & $+0.0079$ & $5/5$ & $5/5$ \\
QSVM                & $-0.0317$ & $-0.0201$ & $1/5$ & $4/5$ \\
VQC                 & $-0.0874$ & $-0.0876$ & $0/5$ & $5/5$ \\
\bottomrule
\end{tabular}
\end{table*}
These results demonstrate that increased retained variance 
from PCA-4 to PCA-8 alone is not a
reliable criterion for selecting the representation dimension. The additional
PCA components are beneficial for the tree ensembles, whereas the same
increase provides no systematic advantage to the RBF--SVM and is detrimental
to the quantum classifiers. The contrast is particularly pronounced for the
VQC, for which the eight-dimensional representation simultaneously increases
the circuit dimension and produces a substantial loss in predictive
performance. PCA-4 therefore remains the appropriate common representation
for the primary classical--quantum scalability comparison, while PCA-8 is
interpreted as a model-dependent representation ablation rather than as a
universally superior feature space.

\section{Conclusions and outlook}
\label{sec:conclusions}

We have presented a controlled comparison of classical and quantum
machine-learning classifiers for Higgs-boson signal--background
discrimination using the ATLAS $H\rightarrow\tau\tau$ challenge dataset.
The study examined the dependence of performance on training-sample
size and representation dimensionality using fixed data partitions,
training-only preprocessing, balanced nested training subsets, model
configurations fixed across training sizes, and matched evaluation
samples. 
The main findings are as follows:

\begin{itemize}

\item
Under the primary PCA-4 representation, the RBF--SVM achieved the highest mean balanced accuracy and event-weighted 
ROC--AUC, while the quantum classifiers remained below the strongest classical benchmark.
 
\item Increasing the training sample size generally improved performance. Among the quantum models, 
the QSVM showed the clearest scaling and narrowed its gap to the RBF--SVM, whereas the VQC showed weaker and 
less consistent behaviour.

\item
The PCA-4 versus PCA-8 comparison showed a model-dependent response to representation dimensionality. 
PCA-8 improved the performance of the tree-based classifiers, whereas the quantum classifiers generally performed 
better with PCA-4, with the VQC showing the degradation under PCA-8. Thus, retaining more input variance does 
not necessarily improve quantum classification performance.

\item AMS-based model rankings did not always agree with those obtained from global classification metrics. 
This indicates that stronger predictive discrimination does not necessarily correspond to greater 
physics-oriented sensitivity, motivating the complementary use of both classification metrics and AMS in high-energy-physics 
applications.

\end{itemize}

The comparatively favourable scaling of the QSVM motivates further
studies with larger training samples and alternative quantum kernels,
feature maps, and encoding schemes. Future work should also
investigate higher-dimensional and physics-informed representations,
improved variational ans\"atze, and more effective optimization
strategies. Evaluations incorporating finite-shot sampling, realistic
device noise, and execution on quantum hardware are needed to
establish whether the observed trends persist under experimentally
relevant conditions.


\bibliography{references}

\end{document}